\documentstyle[twoside,psfig]{article}

\catcode`\@=11
\long\def\@makefntext#1{
\protect\noindent \hbox to 3.2pt {\hskip-.9pt  
$^{{\eightrm\@thefnmark}}$\hfil}#1\hfill}		%CAN BE USED 

\def\@makefnmark{\hbox to 0pt{$^{\@thefnmark}$\hss}}	%ORIGINAL 
	
\def\ps@myheadings{\let\@mkboth\@gobbletwo
\def\@oddhead{\hbox{}
\rightmark\hfil\eightrm\thepage}   
\def\@oddfoot{}\def\@evenhead{\eightrm\thepage\hfil
\leftmark\hbox{}}\def\@evenfoot{}
\def\sectionmark##1{}\def\subsectionmark##1{}}

\oddsidemargin=\evensidemargin
\newcounter{sectionc}\newcounter{subsectionc}\newcounter{subsubsectionc}
\renewcommand{\section}[1] {\vspace{12pt}\addtocounter{sectionc}{1} 
\setcounter{subsectionc}{0}\setcounter{subsubsectionc}{0}\noindent 
	{\tenbf\thesectionc. #1}\par\vspace{5pt}}
\renewcommand{\subsection}[1] {\vspace{12pt}\addtocounter{subsectionc}{1} 
	\setcounter{subsubsectionc}{0}\noindent 
	{\bf\thesectionc.\thesubsectionc. {\kern1pt \bfit #1}}\par\vspace{5pt}}
\renewcommand{\subsubsection}[1] {\vspace{12pt}\addtocounter{subsubsectionc}{1}
	\noindent{\tenrm\thesectionc.\thesubsectionc.\thesubsubsectionc.
	{\kern1pt \tenit #1}}\par\vspace{5pt}}
\newcommand{\nonumsection}[1] {\vspace{12pt}\noindent{\tenbf #1}
	\par\vspace{5pt}}

\newcounter{appendixc}
\newcounter{subappendixc}[appendixc]
\newcounter{subsubappendixc}[subappendixc]
\renewcommand{\thesubappendixc}{\Alph{appendixc}.\arabic{subappendixc}}
\renewcommand{\thesubsubappendixc}
	{\Alph{appendixc}.\arabic{subappendixc}.\arabic{subsubappendixc}}

\renewcommand{\appendix}[1] {\vspace{12pt}
        \refstepcounter{appendixc}
        \setcounter{figure}{0}
        \setcounter{table}{0}
        \setcounter{lemma}{0}
        \setcounter{theorem}{0}
        \setcounter{corollary}{0}
        \setcounter{definition}{0}
        \setcounter{equation}{0}
        \renewcommand{\thefigure}{\Alph{appendixc}.\arabic{figure}}
        \renewcommand{\thetable}{\Alph{appendixc}.\arabic{table}}
        \renewcommand{\theappendixc}{\Alph{appendixc}}
        \renewcommand{\thelemma}{\Alph{appendixc}.\arabic{lemma}}
        \renewcommand{\thetheorem}{\Alph{appendixc}.\arabic{theorem}}
        \renewcommand{\thedefinition}{\Alph{appendixc}.\arabic{definition}}
        \renewcommand{\thecorollary}{\Alph{appendixc}.\arabic{corollary}}
        \renewcommand{\theequation}{\Alph{appendixc}.\arabic{equation}}
        \noindent{\tenbf Appendix \theappendixc #1}\par\vspace{5pt}}
\newcommand{\subappendix}[1] {\vspace{12pt}
        \refstepcounter{subappendixc}
        \noindent{\bf Appendix \thesubappendixc. {\kern1pt \bfit #1}}
	\par\vspace{5pt}}
\newcommand{\subsubappendix}[1] {\vspace{12pt}
        \refstepcounter{subsubappendixc}
        \noindent{\rm Appendix \thesubsubappendixc. {\kern1pt \tenit #1}}
	\par\vspace{5pt}}

\newcommand{\textlineskip}{\baselineskip=13pt}
\newcommand{\smalllineskip}{\baselineskip=10pt}

\def\eightcirc{
\begin{picture}(0,0)
\put(4.4,1.8){\circle{6.5}}
\end{picture}}
\def\eightcopyright{\eightcirc\kern2.7pt\hbox{\eightrm c}} 

\newcommand{\copyrightheading}[1]
	{\vspace*{-2.5cm}\smalllineskip{\flushleft
	{\footnotesize International Journal of Modern Physics B, #1}\\
	{\footnotesize $\eightcopyright$\, World Scientific Publishing
	 Company}\\
	 }}

\newcommand{\publisher}[2]{{\begin{center}\footnotesize\smalllineskip 
	Received #1\\
	Revised #2
	\end{center}
	}}

\def\abstracts#1#2#3{{
	\centering{\begin{minipage}{4.5in}\baselineskip=10pt\footnotesize
	\parindent=0pt #1\par 
	\parindent=15pt #2\par
	\parindent=15pt #3
	\end{minipage}}\par}} 

\renewenvironment{thebibliography}[1]			%ALL CHANGES DD 13/3/92
	{\frenchspacing
	 \ninerm\baselineskip=11pt
	 \begin{list}{\arabic{enumi}.}
	{\usecounter{enumi}\setlength{\parsep}{0pt}
	 \setlength{\leftmargin 12.7pt}{\rightmargin 0pt} %FOR 1--9 ITEMS
	 \setlength{\itemsep}{0pt} \settowidth
	{\labelwidth}{#1.}\sloppy}}{\end{list}}

\newcounter{itemlistc}
\newcounter{romanlistc}
\newcounter{alphlistc}
\newcounter{arabiclistc}

\newenvironment{romanlist}
	{\setcounter{romanlistc}{0}
	 \begin{list}{$($\roman{romanlistc}$)$}
	{\usecounter{romanlistc}
	 \setlength{\parsep}{0pt}
	 \setlength{\itemsep}{0pt}}}{\end{list}}

\newcommand{\fcaption}[1]{
        \refstepcounter{figure}
        \setbox\@tempboxa = \hbox{\footnotesize Fig.~\thefigure. #1}
        \ifdim \wd\@tempboxa > 5in
           {\begin{center}
        \parbox{5in}{\footnotesize\smalllineskip Fig.~\thefigure. #1}
            \end{center}}
        \else
             {\begin{center}
             {\footnotesize Fig.~\thefigure. #1}
              \end{center}}
        \fi}

\newcommand{\tcaption}[1]{
        \refstepcounter{table}
        \setbox\@tempboxa = \hbox{\footnotesize Table~\thetable. #1}
        \ifdim \wd\@tempboxa > 5in
           {\begin{center}
        \parbox{5in}{\footnotesize\smalllineskip Table~\thetable. #1}
            \end{center}}
        \else
             {\begin{center}
             {\footnotesize Table~\thetable. #1}
              \end{center}}
        \fi}

\def\@citex[#1]#2{\if@filesw\immediate\write\@auxout
	{\string\citation{#2}}\fi
\def\@citea{}\@cite{\@for\@citeb:=#2\do
	{\@citea\def\@citea{,}\@ifundefined
	{b@\@citeb}{{\bf ?}\@warning
	{Citation `\@citeb' on page \thepage \space undefined}}
	{\csname b@\@citeb\endcsname}}}{#1}}

\newif\if@cghi
\def\cite{\@cghitrue\@ifnextchar [{\@tempswatrue
	\@citex}{\@tempswafalse\@citex[]}}
\def\citelow{\@cghifalse\@ifnextchar [{\@tempswatrue
	\@citex}{\@tempswafalse\@citex[]}}
\def\@cite#1#2{{$\null^{#1}$\if@tempswa\typeout
	{IJCGA warning: optional citation argument 
	ignored: `#2'} \fi}}

\def\pmb#1{\setbox0=\hbox{#1}
	\kern-.025em\copy0\kern-\wd0
	\kern.05em\copy0\kern-\wd0
	\kern-.025em\raise.0433em\box0}

\def\fnt#1#2{\footnotetext{\kern-.3em
	{$^{\mbox{\scriptsize #1}}$}{#2}}}

\def\fpage#1{\begingroup
\voffset=.3in
\thispagestyle{empty}\begin{table}[b]\centerline{\footnotesize #1}
	\end{table}\endgroup}

\def\runninghead#1#2{\pagestyle{myheadings}
\markboth{{\protect\footnotesize\it{\quad #1}}\hfill}
{\hfill{\protect\footnotesize\it{#2\quad}}}}
\font\tenrm=cmr10
\font\tenit=cmti10 
\font\tenbf=cmbx10
\font\bfit=cmbxti10 at 10pt
\font\ninerm=cmr9
\font\nineit=cmti9
\font\ninebf=cmbx9
\font\eightrm=cmr8

\def\qed{\hbox{${\vcenter{\vbox{			%HOLLOW SQUARE
   \hrule height 0.4pt\hbox{\vrule width 0.4pt height 6pt
   \kern5pt\vrule width 0.4pt}\hrule height 0.4pt}}}$}}

\def\bsc{{\sc a\kern-6.4pt\sc a\kern-6.4pt\sc a}}	%LATEX LOGO
\def\bflatex{\bf L\kern-.30em\raise.3ex\hbox{\bsc}\kern-.14em 
T\kern-.1667em\lower.7ex\hbox{E}\kern-.125em X} 

\begin{document}

\runninghead{C. Grimaldi et al.} 
{Failure of the Migdal-Eliashberg Theory
of Superconductivity in Rb$_3$C$_{60}$}

\normalsize\textlineskip
\thispagestyle{empty}
\setcounter{page}{1}

\copyrightheading{}			%{Vol. 0, No. 0 (1993) 000---000}

\vspace*{0.88truein}

\fpage{1}
\centerline{\bf FAILURE OF THE MIGDAL-ELIASHBERG THEORY}
\vspace*{0.035truein}
\centerline{\bf OF SUPERCONDUCTIVITY IN Rb$_3$C$_{60}$.}
\vspace*{0.37truein}
\centerline{\footnotesize C. GRIMALDI}
\vspace*{0.015truein}
\centerline{\footnotesize\it D\'epartement de Microtechnique-IPM,
\'Ecole Polytechnique F\'ed\'erale de Lausanne}
\baselineskip=10pt
\centerline{\footnotesize\it Lausanne, CH-1015,
Switzerland} 
\vspace*{10pt}
\centerline{\footnotesize E. CAPPELLUTI, L. PIETRONERO}
\vspace*{0.015truein}
\centerline{\footnotesize\it INFM-Unit\'a di Roma 1
and Department of Physics}
\baselineskip=10pt
\centerline{\footnotesize\it University of Rome ``La Sapienza", P.le A. Moro 2}
\baselineskip=10pt
\centerline{\footnotesize\it Rome, I-00185, Italy}
\vspace*{10pt}
\centerline{\footnotesize S. STR\"ASSLER}
\vspace*{0.015truein}
\centerline{\footnotesize\it D\'epartement de Microtechnique-IPM,
\'Ecole Polytechnique F\'ed\'erale de Lausanne}
\baselineskip=10pt
\centerline{\footnotesize\it Lausanne, CH-1015,
Switzerland} 
\vspace*{0.225truein}
\publisher{(received date)}{(revised date)}

\vspace*{0.21truein}
\abstracts{We discuss the compatibility of the most accurate experimental
data with the ordinary Migdal-Eliashberg theory of superconductivity
in the fullerene compound Rb$_3$C$_{60}$.
By using different model phonon spectra we conclude that the
experimental data can be fitted only by invoking an electron-phonon
coupling of order $\lambda\simeq 3$. This exceedingly high value is unphysical
and it is not consistent with the basic assumptions of the Migdal-Eliashberg
theory. On the contrary, by relaxing the adiabatic hypothesis on which the
Migdal-Eliashberg theory rests, the experimental data can be fitted by
much more realistic values of $\lambda$. This generalized theory predicts
also characteristic features absent in the Migdal-Eliashberg framework
which can be experimentally tested.}{}{}

%\vspace*{10pt}
%\keywords{The contents of the keywords}

%\textlineskip			%) USE THIS MEASUREMENT WHEN THERE IS
%\vspace*{12pt}			%) NO SECTION HEADING

\vspace*{1pt}\textlineskip	%) USE THIS MEASUREMENT WHEN THERE IS
\section{Introduction}	%) A SECTION HEADING
\vspace*{-0.5pt}
\noindent
The critical temperatures of the fullerene compounds A$_3$C$_{60}$
(A=K, Rb, Cs, etc.) are the highest
among the organic superconductors and compete in magnitude with those 
of some high-$T_{\rm c}$ superconductors of the cuprate family.\cite{gunny1}
Moreover, if we compare $T_{\rm c}=0.2$ K of the graphite intercalated
compound K$_8$C with $T_{\rm c}=19$ K of K$_3$C$_{60}$ or $T_{\rm c}=30$ K
of Rb$_3$C$_{60}$, it becomes clear that fullerene compounds are
genuine high-$T_{\rm c}$ superconductors.
In addition, as illustrated by the famous Uemura's plot, the fullerene 
compounds, the high-$T_{\rm c}$ cuprates,
the heavy fermions, etc. belong to the same class of anomalous
superconductors characterized by carrier concentrations two or three
orders of magnitude smaller than those of conventional superconductors.\cite{uemura}

Despite of these evidences, it has become a common habit to
regard the fullerene compounds as ordinary superconductors and to
describe them in the framework of the usual Migdal-Eliashberg (ME) theory
of superconductivity, just like the conventional low-temperature
superconductors Hg or Pb.
Such an attitude has been certainly sustained by the relatively
ordinary phenomenology of the normal and superconducting states compared
to that of high-$T_{\rm c}$ cuprates.
Among other things, the fullerene compounds lack in fact 
normal state pseudogaps, have an order parameter of $s$-wave 
symmetry and a sizeable isotope effect for the maximum $T_{\rm c}$. 
This point of view has led to interpret the high values of $T_{\rm c}$ in
fullerene compounds in terms of a strong electron-phonon ($e$-ph)
coupling $\lambda$ generated mainly by intra-molecular phonon modes.

Such an interpretation is actually a quite heavy
over-simplification of the problem and inevitably leads to the odd
conclusion  that, although these materials have extremely low carrier densities, 
may show transitions to ferromagnetic states, are close to a Mott-Hubbard
transition, they are nevertheless the best ME superconductors known 
so far. This situation somehow resembles the one criticised by Anderson and Yu 
in connection with the A15 compounds and the attempts
to interpret their superconducting states in terms of ordinary ME theory.\cite{anderson}

Here, we present evidences against the ME picture commonly advocated for
the fullerene compounds. In particular, we show that the most accurate available
experimental data on Rb$_3$C$_{60}$ are clearly inconsistent with the
ME theory of superconductivity. Moreover, we show also how all
theoretical calculations of the $e$-ph interaction reported
so far lead to results in contradiction with the adiabatic hypothesis which is
at the basis of the ME framework. On the contrary, the relaxation of the adiabatic
hypothesis leads to a more natural interpretation of the experimental data and
defines a theory of nonadiabatic superconductivity which is more promising
than the ME scenario to understand the fullerene compounds.

%\textheight=7.8truein
%\setcounter{footnote}{0}
%\renewcommand{\thefootnote}{\alph{footnote}}

\section{Breakdown of the Migdal-Eliashberg Theory}
\noindent
In the present discussion of the compatibility of the experimental
data with the ME theory of superconductivity we refer solely to 
Rb$_3$C$_{60}$. The reason is that only for this material sufficiently
accurate data have become available. In fact, from both tunneling and
infrared measurements, the ratio gap-$T_{\rm c}$ in Rb$_3$C$_{60}$
has been measured to be $2\Delta/T_{\rm c}=4.2\pm 0.2$,\cite{koller} 
and the most
accurate measurements of the carbon isotope effect on $T_{\rm c}$ has
led to $\alpha_{\rm C}=0.21$, where $\alpha_{\rm C}=-d\log T_{\rm c}/
d\log M_{\rm C}$ is the carbon isotope coefficient.\cite{fuhrer} As we show below, 
together with $T_{\rm c}=30$ K this set of data permits to test the 
ME theory for Rb$_3$C$_{60}$.

Let us begin by considering the following standard strong-coupling formulas
derived form the ME equations:\cite{carbotte}
\begin{equation}
\label{tcmcmillan}
T_{\rm c}=\frac{\omega_{\rm ln}}{1.2}
\exp\left[-\frac{1.04(1+\lambda)}{\lambda-\mu^*(1+0.62\lambda)}\right] ,
\end{equation}
\begin{equation}
\label{alpha}
\alpha_{\rm C}=\frac{1}{2}\left[1-
\frac{1.04(1+\lambda)(1+0.62\lambda)\mu^{*2}}{[\lambda-\mu^*(1+0.62\lambda)]^2}
\right] ,
\end{equation}
\begin{equation}
\label{delta-tc}
\frac{2\Delta}{T_{\rm c}}=3.53\left[1+12.5\left(\frac{T_{\rm c}}
{\omega_{\rm ln}}\right)^2\ln
\left(\frac{\omega_{\rm ln}}{2T_{\rm c}}\right)\right] ,
\end{equation}
where $\lambda=2\int\alpha^2\!F(\omega)d\omega/\omega$ is the 
$e$-ph coupling, $\ln\omega_{\rm ln}=(2/\lambda)
\int\ln\omega \alpha^2\!F(\omega)d\omega/\omega$ is the relevant phonon
frequency and $\alpha^2\!F(\omega)$ is the $e$-ph spectral
fuction (also known as Eliashberg function). In equations (\ref{tcmcmillan})
and (\ref{alpha}) $\mu^*$ is the Coulomb repulsive pseudopotential.
Inserting the above reported experimental values of $T_{\rm c}$, $\alpha_{\rm C}$ and
$2\Delta/T_{\rm c}$ in the left-hand sides of equations 
(\ref{tcmcmillan})-(\ref{delta-tc}) completely determines the values
of the three unknown microscopic parameters $\lambda$, $\mu^*$ and
$\omega_{\rm ln}$ (for the moment we neglect the error bars of $\Delta$ and
assume therefore $2\Delta/T_{\rm c}=4.2)$. 
The unique solution is given by $\omega_{\rm ln}=313$ K,
$\mu^*=0.43$ and $\lambda=3.6$. Although the obtained values of
$\omega_{\rm ln}$ and $\mu^*$ lie within a reasonable range of validity, the
extremely large value $\lambda=3.6$ is completely unrealistic.
There are mainly two reasons for this conclusion.
First, such a strong $e$-ph interaction is expected to be source
of lattice instabilities preventing the system to become superconductive.
In fact, although the maximum allowed value of $\lambda$ compatible with 
superconductivity is not precisely known, it is generally believed that
$\lambda\sim 1.5$ represents a reasonable upper limit.\cite{anderson}
Instead this limiting value is largely surpassed by the ME solution $\lambda=3.6$.

A second reason which makes $\lambda=3.6$ incompatible with the ME theory
is given by observation that the half-filled electron conduction bandwidth
of Rb$_3$C$_{60}$ (as for the other fullerene compounds) 
is of order $W=0.5$ eV.\cite{gunny1}
In this situation, the $e$-ph vertex corrections which are completely neglected 
in the ME theory become important. In fact, 
the order of magnitude of the vertex corrections is roughly 
$\lambda \omega_{\rm ph}/E_{\rm F}$, where $\omega_{\rm ph}$ is an averaged 
phonon frequency and $E_{\rm F}$ is the Fermi energy.\cite{migdal} For conventional
superconductors $\lambda \omega_{\rm ph}/E_{\rm F}$ is of order 
$10^{-3}$-$10^{-4}$ and the $e$-ph vertex corrections can be safely neglected.
Instead by using $\lambda=3.6$, $\omega_{\rm ph}\sim\omega_{\rm ln}=313$ K
and $E_{\rm F}=W/2\simeq 0.25$ eV,
we obtain $\lambda \omega_{\rm ph}/E_{\rm F}\simeq 0.35$. This result reveals
the inconsistency of the ME theory when applied to Rb$_3$C$_{60}$ since
the vertex corrections, which should be negligible in the ME framework,
turn out to be instead important.

The above analysis is confirmed also
when, in place of the strong-coupling formulas 
(\ref{tcmcmillan})-(\ref{delta-tc}), we solve numerically the ME equations
to fit the experimental data $T_{\rm c}=30$ K, $\alpha_{\rm C}=0.21$, 
and $2\Delta/T_{\rm c}=4.2\pm 0.2$. In Fig.1 we report the calculated
values of $\lambda$ for a $e$-ph spectral function $\alpha^2\!F(\omega)$
schematized with a rectangle centered at $\omega_0$ and width $\Delta\omega_0$.
For the whole range of $\Delta\omega_0/\omega_0$ values compatible
with the vibrational frequencies of the fullerene molecule we obtain
$\lambda\sim 3$, in agreement with the analysis based on the strong-coupling
formulas. Also the calculated values $\omega_{\rm ln}\sim 350$ K and
$\mu^*\sim 0.35$ are close to the values obtained above.

\begin{figure}[t]
\vspace*{13pt}
\centerline{\psfig{figure=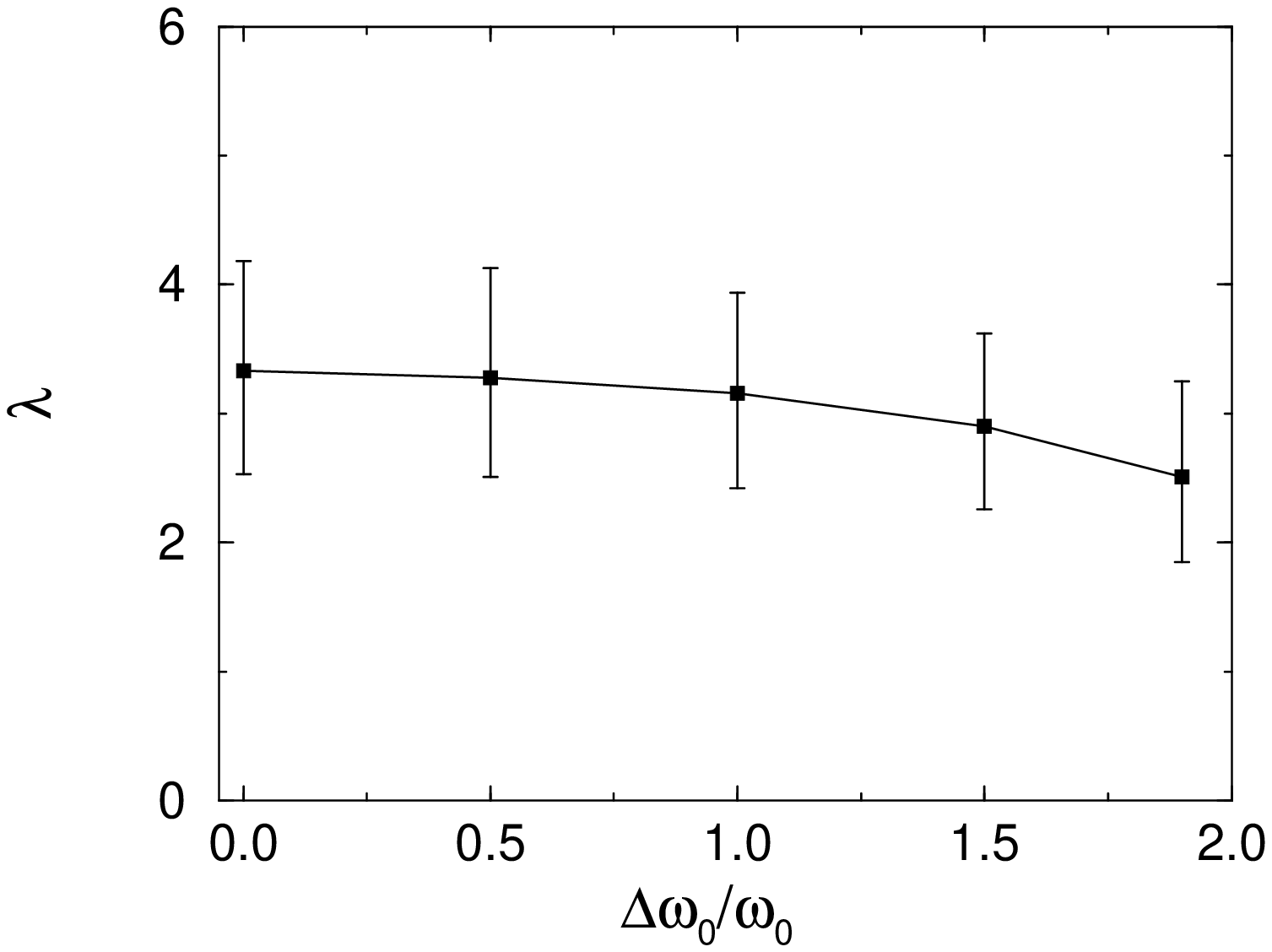,width=6cm}
\psfig{figure=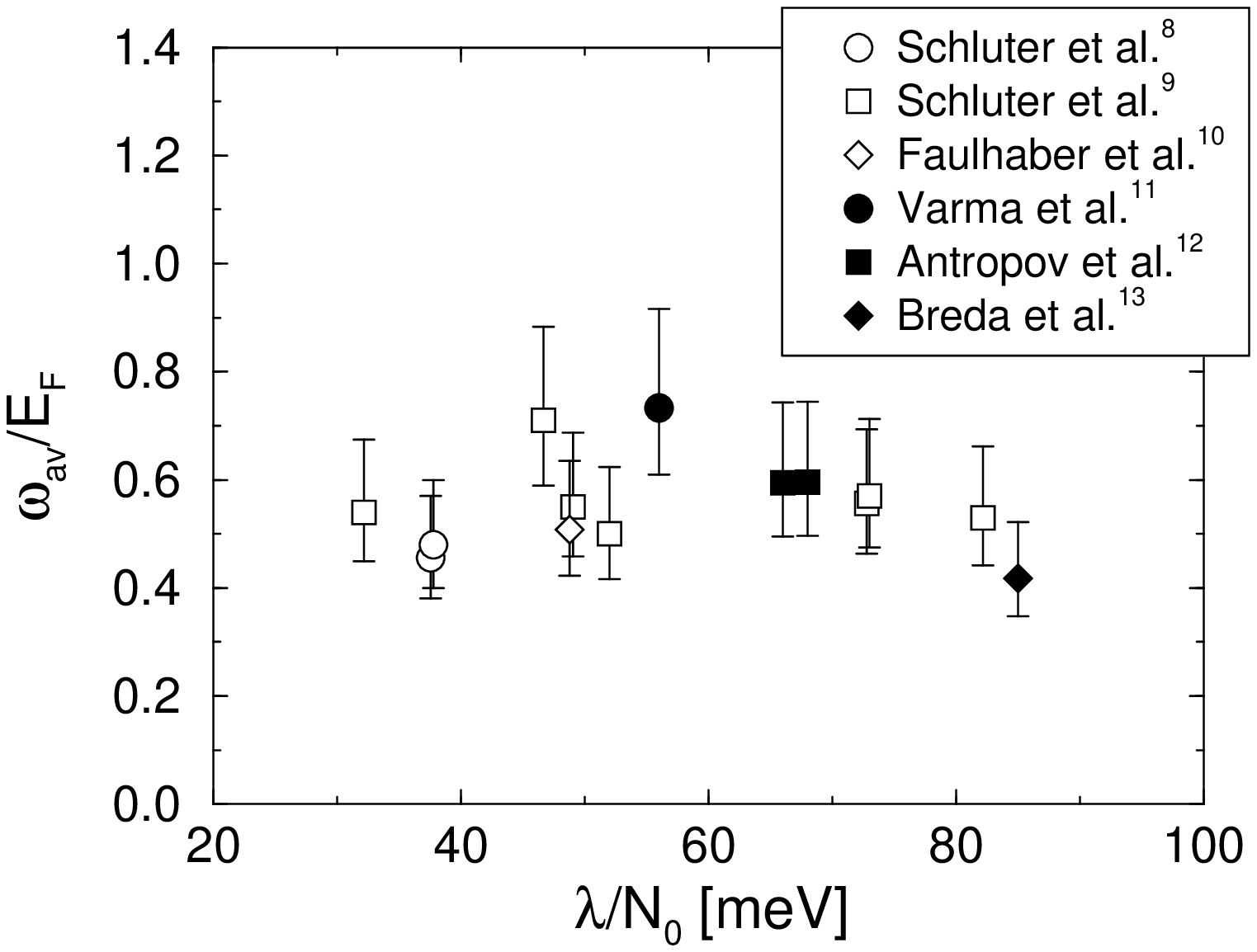,width=6cm}}
%\centerline{\psfig{figure=omega_EF.eps,width=5cm}}
\vspace*{13pt}
\fcaption{Electron-phonon coupling constant calculated from the solution
of the ME equations for an Eliashberg function of rectangular form centered at $\omega_0$
and having width $\Delta\omega_0$. The ME equations have been solved by requiring
to fit the experimental results $T_{\rm c}=30$ K, $\alpha_{\rm C}=0.21$, 
and $2\Delta/T_{\rm c}=4.2\pm 0.2$.}
\fcaption{Adiabatic parameter $\omega_{\rm ph}/E_{\rm F}$ 
extracted from various calculations of the $e$-ph interaction in fullerenes.
The average phonon frequency is calculated from 
$\omega_{\rm ph}=(1/\lambda)\sum_i\lambda_i\omega_i$, where the index $i$ runs over
the eight H$_{\rm g}$ phonon modes with $e$-ph couplings and frequencies 
$\lambda_i$ and $\omega_i$, respectively. 
The Fermi energy is $E_{\rm F}=0.25\pm 0.05$ eV.}
\end{figure}

The above numerical evaluations of the ME equations confirm and strengthen therefore
our conclusion that ME theory fails to describe superconductivity of
Rb$_3$C$_{60}$. This situation is also supported by the various calculations
of the $e$-ph interaction in fullerenes reported so far. In Fig.2 we show a collection
of data taken from various theoretical calculations of the $e$-ph interaction $V=\lambda/N_0$,
$N_0$ being the electron density of states per spin. The data refer to various
calculations schemes including tight-binding, LDA,  ab-initio etc. which estimate the
coupling of the $t_{\rm 1u}$ electrons to the eight H$_{\rm g}$ intra-molecular 
phonon modes.\cite{schluter1,schluter2,faulhaber,varma,antropov,breda}
As it is apparent from the values reported in the abscissa, there is a large uncertainty
in the value of $\lambda/N_0$ which in fact results equally distributed between 
$32$ meV and $85$ meV, {\it i.e.} a variation of more than $100$ \%.
However, despite of the serious disagreement on the value of $\lambda/N_0$,
all these calculations agree in estimating the adiabatic parameter $\omega_{\rm ph}/E_{\rm F}$
to be larger than $0.4$. In this case, Migdal's theorem breaks down and the whole ME
framework is invalidated. The reason for such high values of $\omega_{\rm ph}/E_{\rm F}$
stems from the fact that, independently of details, the H$_{\rm g}$ phonons have 
energies ranging from $30$ meV to $200$ meV while the conduction $t_{\rm 1u}$ 
electron band has a width of only $W=0.4-0.6$ eV,\cite{gunny1} 
so that $E_{\rm F}=W/2=0.25\pm 0.05 $ eV
and the electrons and phonons can have comparable energies.

\section{Beyond Migdal's Limit}
\noindent
The failure of the ME theory of fullerene compounds revealed from the analysis
of both experimental (Fig.1) and theoretical (Fig.2) data suggests that the $e$-ph problem
in these materials must be approached from a different perspective.
From the above discussion, it is clear that the most natural starting point is to relax 
the adiabatic hypothesis on which the ME framework rests and consequently to 
generalize the theory to the nonadiabatic regime of the $e$-ph interation.
Within a perturbative scheme, this generalization means that nonadiabatic contributions
such as $e$-ph vertex corrections must be included in the evaluation of both normal
and superconducting state properties. A detailed derivation of such a generalized theory
beyond Migdal's limit has already been reported elsewhere.\cite{grima1} 
Here it is sufficient to remark
that the nonadiabatic corrections are strongly dependent on the exchanged phonon 
frequency $\omega$ and momentum transfer $q$ in such a way that for 
$v_{\rm F}q/\omega<1$ ($v_{\rm F}q/\omega >1$) the resulting $e$-ph effective scattering
is enhanced (diminished). As a consequence, the critical temperature $T_{\rm c}$ can
be amplified by the nonadiabatic corrections if the $e$-ph scattering is mainly via small
momentum transfer,\cite{grima1} as it is expected for strongly correlated systems and the weak screening
due to the low charge carrier concentration.\cite{grilli}

Having a theory of nonadiabatic superconductivity available, let us now apply it to the
situation of Rb$_3$C$_{60}$. The evaluation of the zero temperature gap $\Delta$
for nonadiabatic superconductors is a quite difficult task. However, there
are various indications that the pseudopotential $\mu^*$ is about $0.3$.\cite{gunny2} 
Let us consider
therefore only the experimental data $T_{\rm c}=30$ K and $\alpha_{\rm C}=0.21$,
and fix the value $\mu^*=0.3$. In this case, for a dispersionless phonon spectrum with 
frequency $\omega_0$,  the ordinary ME equations lead to a unique solution with
$\lambda=3.5$ and $\omega_0=300$ K, very close therefore to the values obtained in
the previous section. However, when we solve the generalized equations which include
the nonadiabatic contributions, we find that for $\mu^*=0.3$ the experimental data
$T_{\rm c}=30$ K and $\alpha_{\rm C}=0.21$ can be fitted by much more reasonable
values of $\lambda$.\cite{emmcapp} In Table 1 we report the results for different values of the
dimensionless momentum transfer $Q_{\rm c}=q_{\rm c}/2k_{\rm F}$. The resulting
$e$-ph coupling $\lambda$ depends only weakly on $Q_{\rm c}$ and it is much lower
than the unrealistically large values obtained by the ME analysis. 
In the last column of Table 1 we report also
the values of $\lambda\omega_0/E_{\rm F}$ ($E_{\rm F}=0.25$ eV) which provide a 
measure of the importance of the nonadiabatic corrections.

\begin{table}[htbp]
\tcaption{Electron-phonon coupling $\lambda$, phonon energy $\omega_0$
and Migdal parameter $\lambda\omega_0/E_{\rm F}$ for three different values of
the momentum transfer $Q_{\rm c}=q_{\rm c}/2k_{\rm F}$ reproducing
the experimental data of Rb$_3$C$_{60}$: $T_{\rm c}=30$ K and $\alpha_{\rm C}=0.21$.
The pseudopotential $\mu^*$ is set equal to $0.3$.}
\centerline{\footnotesize\smalllineskip
\begin{tabular}{c c c c}\\
\hline
$Q_{\rm c}$ & $\lambda$ & $\omega_0$ [K] & $\lambda\omega_0/E_{\rm F}$ \\
\hline
0.1 & 0.46 & 1519 & 0.24\\
0.3 & 0.55 & 2340 & 0.44\\
0.5 & 0.76 &1787 & 0.47\\
\hline\\
\end{tabular}}
\end{table}

\section{Discussion and Conclusions}
\noindent
The quite promising results on the experimental data of Rb$_3$C$_{60}$ are solid
arguments in favour of the nonadiabatic nature of the $e$-ph interaction in the
fullerene compouns. Moreover, the nonadiabatic scenario is liable to be experimentally 
tested in different ways. Below we list the features we have identified to be characteristic
of a nonadiabatic $e$-ph interaction and that cannot be displayed by a ME material:
\begin{romanlist}
\item The critical temperature $T_{\rm c}$ is strongly suppressed by non-magnetic
impurity scattering and the isotope coefficient acquires an anomalous dependence
on disorder.\cite{scattoni}
\item The {\it normal state} electron effective mass $m^*$ becomes ion-mass dependent 
leading to a negative isotope coefficient of $m^*$.\cite{grima2}
\item The {\it normal state} Pauli susceptibility $\chi$ acquires a dependence on the 
$e$-ph scattering. For $\omega_0/E_{\rm F}\neq 0$, the $e$-ph interaction leads
to a decreasing of $\chi$ with respect to the adiabatic limit 
$\omega_0/E_{\rm F}\rightarrow 0$. Moreover, in analogy with $m^*$, $\chi$ has a nonzero
negative isotope effect.\cite{grima3}
\end{romanlist}
We note that the effect described in point (i) has been already observed in 
K$_3$C$_{60}$.\cite{watson} We conclude by stressing that
experimental verifications of the predictions listed above is of primary importance to
settle the degree of nonadiabaticity of the $e$-ph interaction in the fullerene compounds.

\nonumsection{References}

\end{document}